# "Birds of a Feather": Does User Homophily Impact Information Diffusion in Social Media?

Munmun De Choudhury,[*] Hari Sundaram,[†]
Ajita John,[‡] Doree Duncan Seligmann,[§]
Aisling Kelliher[¶]

**Abstract**

This article investigates the impact of user homophily on the social process of information diffusion in online social media. Over several decades, social scientists have been interested in the idea that *similarity breeds connection*—precisely known as "homophily". "Homophily", has been extensively studied in the social sciences and refers to the idea that users in a social system tend to bond more with ones who are "similar" to them than to ones who are dissimilar. The key observation is that homophily *structures* the ego-networks of individuals and impacts their communication behavior. It is therefore likely to effect the mechanisms in which information propagates among them. To this effect, we investigate the interplay between homophily along diverse user attributes and the information diffusion process on social media.

Our approach has three steps. First we extract several diffusion characteristics along categories such as user-based (volume, number of seeds), topology-based (reach, spread) and time (rate)—corresponding to the baseline social graph as well as graphs filtered on different user attributes (e.g. location, activity behavior). Second, we propose a Dynamic Bayesian Network based framework to predict diffusion characteristics at a future time slice. Third, the impact of attribute homophily is quantified by the ability of the predicted characteristics in explaining actual diffusion, and external temporal variables, including trends in search and news. Experimental results on a large Twitter dataset are promising and demonstrate that the choice of the homophilous attribute can impact the prediction of information diffusion, given a specific metric and a topic. In most cases, attribute homophily is able to explain the actual diffusion and external trends by $\sim 15-25\%$ over cases when homophily is not considered. Our method also outperforms baseline techniques in predicting diffusion characteristics subject to homophily, by $\sim 13-50\%$.

[*]School of Computing, Informatics & Decision Systems Engineering, Arizona State University, Tempe, Arizona, USA. (`munmun@asu.edu`).

[†]School of Arts, Media & Engineering, Arizona State University, Tempe, Arizona, USA. (`hari.sundaram@asu.edu`).

[‡]Collaborative Applications Research, Avaya Labs Research, Basking Ridge, New Jersey, USA. (`ajita@avaya.com`).

[§]Collaborative Applications Research, Avaya Labs Research, Basking Ridge, New Jersey, USA. (`doree@avaya.com`).

[¶]School of Arts, Media & Engineering, Arizona State University, Tempe, Arizona, USA. (`aisling.kelliher@asu.edu`).

# 1 Introduction

The central goal in this article is to investigate the relationship between homophily among users and the social process of information diffusion. The "homophily" principle: the idea that users in a social system tend to bond more with ones who are "similar" to them than ones who are dissimilar, has been a popular area of investigation to social scientists over several decades [7, 23, 22]. Predominantly ethnographic and cross-sectional in nature, such studies have revealed that homophily *structures* networks—people's ego-centric social networks are often homogeneous with regard to diverse social, demographic, behavioral, and intra-personal characteristics [22] or revolve around social foci such as co-location or commonly situated activities [12]. As a consequence of interaction in these homogeneous constructs, individuals tend to become interpersonally tied to each other. Hence, the existence of such homogeneity, i.e. homophily is likely to impact the information these individuals receive and propagate, the communication activities they engage in, and the social roles they form.

The advent of the social web in the past few years has given leeway to a broad rubric of platforms such as Facebook, YouTube, Digg and Twitter that have bolstered these sociological findings. These networks facilitate the sharing and propagation of information among members of their networks.Moreover, these social sites often exhibit evidences of the presence of homophilous relationships and subsequently their impact on associated social phenomena at *very large scales*: such as group evolution and information diffusion. For example, the popular social networking site Facebook allows users to engage in community activities via homophilous relationships involving common organizational affiliations. Whereas the video-sharing service YouTube features extensive communication activity (in the form of nested commentary) among users who share similar interests in a video. Finally, on the fast-growing social media Twitter, several topics such as '#Elections2008', '#MichaelJackson', 'Global Warming' etc have historically featured extensive postings (also known as "tweets") due to the common interests of large sets of users in politics, music and environmental issues respectively.

These networks, while diverse in terms of their affordances (i.e. what they allow users to do), share some common features. First, there exists a social action (e.g. posting a tweet on Twitter) within a shared social space (i.e. the action can be observed by all members of the users' contact network), that facilitates a social process (e.g. diffusion of information). Second, these networks expose attributes including location, time of activity and gender to other users. Finally, these networks also reveal these users attributes as well as the communication, to third party users (via the API tools). We therefore conjecture that on online social media, there is likely to be a strong interplay between the social process, e.g. information diffusion, and the presence of homophilous relationships along shared attributes of users, e.g. their location, information roles or their activity behavior. However, note that most social processes, particularly information diffusion, although have been explored rather extensively [16, 15, 21], but the role of homophily has not been investigated substantially.

There are several reasons why understanding the impact of homophily on diffusion can be deemed to be of value. We consider some example applications. Today, due to the plethora of diverse retail products available online to customers, advertising is moving from the traditional "word-of-mouth" model to models that exploit interactions among individuals on social networks. To this effect, previously, some studies have provided useful insights that social relationships impact the adoption of innovations and products [16]. Moreover there has been theoretical and empirical evidence in prior work [35] that indicates that individuals have been able to transmit information through a network (via messages) in a sufficiently small number of steps, due to homophily along recognizable personal identities. Hence, apart from considering interactions, a viral marketer attempting to advertise a new product could benefit from considering specific sets of users on a social space who are homophilous with respect to their interest in similar products or features. In another example in this context, suppose a crisis mitigation team is attempting to estimate vulnerability and risk of people subject to a natural calamity. In this case it might be useful for the team to focus on the set of people from that particular location and utilize their shared information content to promote rehabilitation. Besides, understanding the impact of homophily on diffusion is likely to have potential in addressing the propagation of medical and technological innovations, cultural bias, in understanding social roles and in distributed social search.

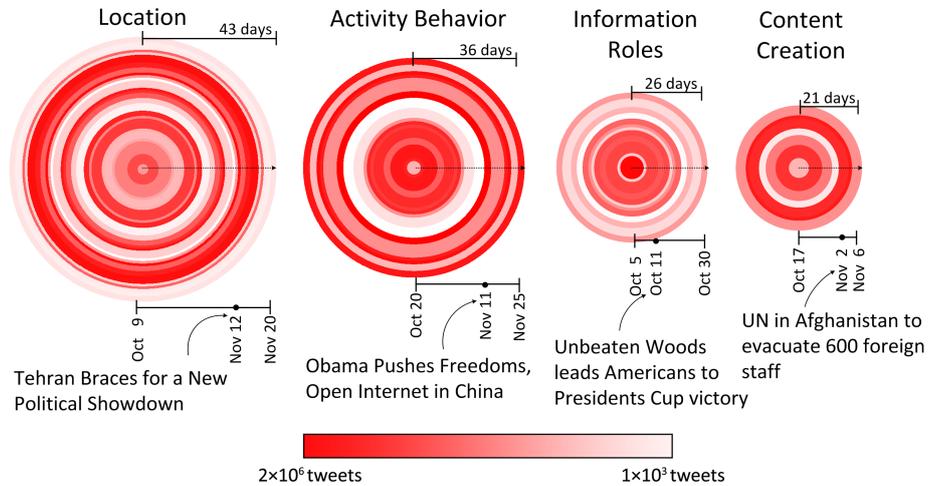

Figure 1: Examples of how attribute homophily impacts diffusion. We visualize diffusion on Twitter on the theme 'Politics' as a "ripple" (of cascades) over time and consider four different attribute social graphs: location, information roles, content creation and activity behavior (ref. section 3.1.1). As noted, the discovered diffusion phenomenon is significantly different for the four attributes—e.g. the time span (size) of the ripple, the volume of users involved (width of the arcs) as well as the extent of tweeting activity (color intensity of each arc).

To this end, our central goal in this article is to investigate the relationship between homophily among users and the social process of diffusion on the social media Twitter.

## 1.1 Motivating Study

We motivate our problem domain through a qualitative study on Twitter data. The study reveals how different attributes affect the diffusion process on a particular theme "Politics" (comprising topics such as 'Obama' and 'Tehran') during Oct-Nov 2009. Figure 1 presents a "ripple" visualization of the diffusion process over a set of social graphs constructed using the attributes—location, information roles, content creation and activity behavior. We describe the visualization as follows. In each ripple defined over a chosen attribute over which homophily is likely to exist among the users, we represent time (in days) by each arc. The color intensity of each arc represents the extent of user activity (in terms of frequency of tweets) on the corresponding day. While the width of the arc indicates the extent of user involvement (in terms of unique users who tweet) on the same day.

The visualization reveals that the choice of the attribute has a huge impact on the discovery of diffusion properties. For example, location seems to yield diffusion ripples over the longest period of time, while content creation the shortest. This implies that the diffusion of information on "Politics" takes place extensively over the location attribute of users, i.e. there exists homophily along the location attribute corresponding to the diffusion process over "Politics". Our conjecture in the explanation of this finding is that "Politics", is highly related to local happenings with respect to sets of users. Hence it is able to quantify the diffusion process the best among all the chosen attributes, in terms of predicting ripples over long periods of time. Whereas the content creation attribute, being primarily reflective of the intrinsic habits of users, is not able to characterize the diffusion process involving external events satisfactorily.

Moreover, it appears that the particular attribute also affects the spatial location of the time periods (or arcs) of high color intensity (i.e. high tweeting activity) in each ripple. On studying the associated news events (http://news.google.com/) as annotated on the Figure, we observe qualitative correlation between the attribute and the topic of the news event. For example, the news on Tehran seems to be of interest to users of certain locations; while the news related to Obama's China visit, being a temporal event, is likely of interest to users with a certain activity behavior over time.

The observation that consideration of homophilies across different attributes results in marked differences in the discovered diffusion process, can be seen across different topics as well. For example, in Figure 2, we show the best and worst performing attributes (in terms of the width or size of the diffusion ripple discovered) for two topics—"Politics" as above, and "Technology-Internet". Note that while the attribute "information roles" performs poorly for the first, it performs extremely well for the latter. This is explained by our intuition that for topics such as technology, there is likely to be strong homophily along the responsive behavior of users, i.e. how they respond to tweets, compared to other attributes such as location.

Hence the above motivating study shows that consideration of homophily along the *right* attribute is extremely important to researchers attempting to predict diffusion processes over online social media. This forms the basic premise of this work.

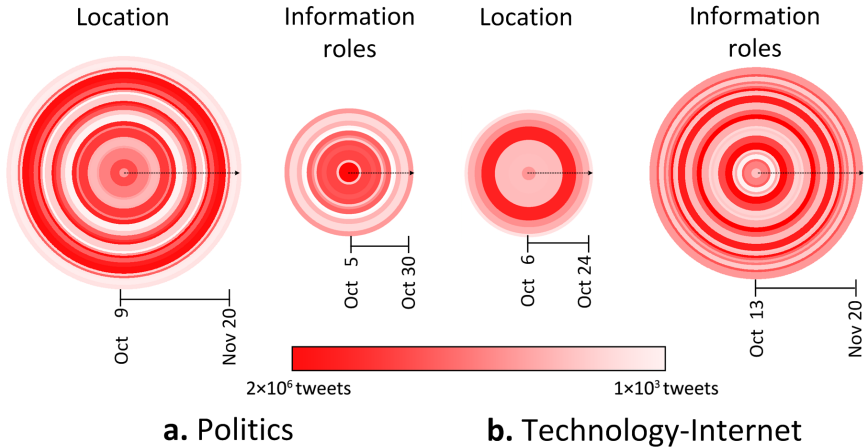

**Figure 2:** Comparative ripple visualization of discovered diffusion processover two topics: "Politics" and "Technology-Internet". The attributes that yield the best and the worst performance, in terms of the size of the diffusion ripple discovered, are shown.

## 1.2 Main Contributions

There are two key contributions in this article:

1. We propose a three step approach to investigate the role played by homophily in predicting diffusion characteristics on a given topic over time. First, we extract diffusion characteristics along different categories, such as user-based (volume, number of seeds), topology-based (reach, spread) and time (rate), corresponding to social graphs defined on different user attributes (e.g. location, activity behavior). Second, we predict the users likely to get involved in the diffusion process at a future time slice based on a Dynamic Bayesian Network based probabilistic framework. Third, we utilize the predicted set of users to determine diffusion characteristics at the future time slice. We quantitatively define distortion metrics to study how the predicted characteristics corresponding to each attribute (i.e. presence of homophily along a certain attribute) can explain the actual characteristics as well as external time-series variables—search and news trends.

2. We demonstrate, based on a large dataset from Twitter, that the choice of different attributes can impact the prediction of diffusion process differently based on the metric used to quantify diffusion and the topic under consideration. In most cases, attribute homophily is able to explain the actual diffusion and external trends by a margin of $\sim 15-25\%$ lower distortion compared to cases when homophily is not considered. Comparison with baseline techniques also reveals that our method outperforms others in predicting diffusion characteristics subject to homophily, by $\sim 13-50\%$.

The rest of this article is organized as follows. We discuss related work in section 2. In section 3, we present our problem formulation. Sections 4, 5 and 6 describe our proposed method. We discuss the experimental results in section 7 and conclude in section 8 with our major contributions.

## 2 Related Work

The role of homophily in the formation and sustenance of social ties and networks has been studied extensively by sociologists since several decades [12, 7, 23, 22, 18]. We present prior research relevant to this work from three different perspectives—attribute homophily in the context of social networks, information diffusion subject to shared social actions, and role of homophily in social contagion.

### 2.1 Attribute Homophily

The role of homophily in the formation and sustenance of social ties and networks have been studied extensively by sociologists since several decades [12, 7, 23, 22, 18]. In the two seminal works [23, 22], McPherson et al study the inter-connectedness between homogeneous composition of groups and emergent homophily, and how user context such as geographic propinquity, kinship and isomorphic structural locations in a network allow the formation of homophilous relations. In [12], the author presents theories to explain the origins of social circles, in terms of social foci that are social, psychological or physical entities around which joint activities of users are organized, e.g. workplaces, families and hangouts. Social foci are considered to be central to how ties are built—consequently how homophily emerges. Fiore et al in [13] investigate the role of homophily in online dating choices made by users.

### 2.2 Information Diffusion

Understanding the diffusion of information in social networks is one of the oldest topics of interest to researchers. The analysis of social information diffusion has been of interest to researchers from various domains ranging from social sciences, epidemiology, physics and economics [35, 16, 15, 5, 21]. There has been prior work on mining and predicting pathways of diffusion in social networks useful for several applications, ranging from recommendation systems, online advertising, user behavior prediction and disease containment [29, 30, 33].

Bass in [4] proposed a network independent method to determine the the rate of diffusion at a certain time, which was based on the rate at a previous time, a coefficient of adoption and a coefficient of incitation in the market, based on word-of-mouth. In an early work [16], Kempe et al propose solution to the optimization problem of selecting the most influential nodes in a social network which could trigger a large cascade of further adoptions. They use sub-modular functions and a greedy strategy to yield approximate solutions that out-perform node-selection heuristics based on the popular notions of degree centrality and distance centrality. In [15] the authors focus on analyzing the text in blog posts and use an epidemic disease propagation model for determining information diffusion. The authors focus on the propagation of topics from one blog to the next in the blogosphere, based on the textual content instead of analysis of the hyperlinking structure. Using this information, they have characterized information diffusion along two dimensions: topics and users.

Adar and Adamic in [1] utilize a novel inference scheme that takes advantage of data describing historical, repeating patterns of 'infection'. They present a visualization system that allows for the graphical tracking of such information flow. In [29], the authors present an early adoption based information flow model useful for recommendation systems. The authors in [26] provide simple models for the onset of epidemic behavior in diseases. Their results derived exact analytic expressions for the percolation threshold on one-dimensional small-world graphs under both site and bond percolation. They have also looked briefly at the case of simultaneous site and bond percolation, in which both susceptibility and transmissibility can take arbitrary values.

In [14], the authors propose a novel model for networks of complex interactions, based on a granular system of mobile agents whose collision dynamics is governed by an efficient event-driven algorithm and generate the links (contacts) between agents. There has also been some prior work [11] where the authors explain how diffusion dynamics in small world networks are affected with having heterogeneous consumers. Stewart et al in [30] propose an algorithm to discover information diffusion paths in the blogosphere for helping online advertising domains. They present a 'frequent pattern mining' based method in which they focus on analyzing blog content and topic extraction. Using sequences of blogs, they discover information diffusion paths which are useful for effective information flow in social networks.

In [33], Wan and Yang define information diffusion to be the phenomenon of document forwarding or transmission between various web sites on the Web. They propose a method for mining information diffusion processes for specific topics on the Web and develop a system called LIDPW to address this problem using matching learning techniques. Saito et al in [28] utilize the independent cascade (IC) model to determine the likelihood for information diffusion episodes, where an episode is defined as a sequence of newly active nodes. Thereafter they present a method for predicting diffusion probabilities by using the popular EM algorithm.

In a recent work, Bakshy et al [3] study how "gestures" make their way through an online community—the social gaming environment called Second Life. Gestures are code snippets that Second Life avatars must acquire in order to make motions such as dancing, waving or chanting. Their empirical studies indicate that individuals who have already declared each other as friends are more prone to getting influenced by each other and subsequently acquiring assets, rather than two individuals arbitrarily apart. In another recent work, Sun, Rosenn, Marlow et al [31] study the diffusion patterns on the Facebook "News Feed" and conclude that in online social media, diffusion dynamics are often triggered by the collision of short chains of information trigger, rather than the preconceived notion in the literature that diffusion occurs due to several long chains generated by a small number of "seeds". Tang et al in [32] propose a topic affinity propagation (TAP) model for modeling social influence in large networks.

## 2.3 Homophily and Social Contagion

A body of prior work also studies the interplay between the homophily principle and social contagion; particularly in the context of romantic and sexual networks [5]. Their primary observation, largely based on cross-sectional studies of populations, is that network structures inferred based on homophily in partner preferences e.g. race, religiosity are able to mimic the observed spread of sexually transmitted diseases. In their review paper [8], Cialdini et al discuss principal driving forces behind spread of social influence and therein the role played by conformity and compliance among individuals, e.g. affiliations, self-categorization. Crandall et al in [9] study the interplay between similarity, emergence of social ties and subsequent social influence on the Wikipedia community. Finally, a recent work [6] also explores the impact of "similarity networks" on the design of online social content aggregation services and recommender systems.

Although motivated by different research questions, the approaches taken in these studies do not provide any comprehensive computational analysis of the impact of attribute homophily on diffusion of content in large-scale social media datasets. Unlike problems involving disease containment in networks, information shared on social platforms such as Twitter are extremely content rich (i.e. diverse topics) and can often be correlated with external events [20]. Moreover the diversity of users in terms of activity, demographics and roles is likely to induce various homophilous social relationships. Hence certain kinds of homophilous ties are likely to be more conducive to the flow of certain types of information. Additionally, specific online communities would presumably be interested in information content that is available to them along a homophilous social dimension, than that dissipated via traditional RSS feeds or retrieval techniques. To the best of our knowledge, this is the first time such interplay between homophily and diffusion is being investigated in depth in the context of online social media.

## 3 Problem Formulation

In this section we present our problem formulation. First, we present the key concepts involved in this article, followed by the problem statement.

### 3.1 Preliminaries

We first present the social graph model used in this article, and then definitions of attribute homophily and diffusion. Finally we introduce a structure called a diffusion series to quantify the diffusion in the social graph over time.

#### 3.1.1 Social Graph Model

We define our social graph model as a directed graph $G(V, E)$[1], such that $V$ is the set of users and $e_{ij} \in E$ if and only if user $u_i$ and $u_j$ are "friends" of each other (bi-directional contacts). Let us further suppose that each user $u_i \in V$ can perform a set of "social actions", $\mathcal{O} = \{O_1, O_2, \ldots\}$, e.g. posting a tweet, uploading a photo on Flickr or writing on somebody's Facebook Wall. Let the users in $V$ also be associated with a set of attributes $\mathcal{A} = \{a_k\}$ (e.g. location or organizational affiliation) that are responsible for homophily. Corresponding to each value $v$ defined over an attribute $a_k \in \mathcal{A}$, we construct a social graph $G(a_k = v)$ such that it consists of the users in $G$ with the particular value of the attribute, while an edge exists between two users in $G(a_k)$ if there is an edge between them in $G$.[2] E.g., for location, we can define sets of social graphs over users from Europe, Asia etc.

---
[1]Henceforth referred to as the baseline social graph $G$.

In this article, our social graph model is based on the social media Twitter. Twitter features a micro-blogging service that allows users to post short content, known as "tweets", often comprising URLs usually encoded via bit.ly, tinyurl, etc. The particular "social action" in this context is the posting of a tweet; also popularly called "tweeting". Users can also "follow" other users; hence if user $u_i$ follows $u_j$, Twitter allows $u_i$ to subscribe to the tweets of $u_j$ via feeds; $u_i$ is then also called a "follower" of $u_j$. Two users are denoted as "friends" on Twitter if they "follow" each other. Note that, in the context of Twitter, using the bi-directional "friend" link is more useful compared to the uni-directional "follow" link because the former is more likely to be robust to spam—a normal user is less likely to follow a spam-like account. Further, for the particular dataset of Twitter, we have considered a set of four attributes associated with the users:

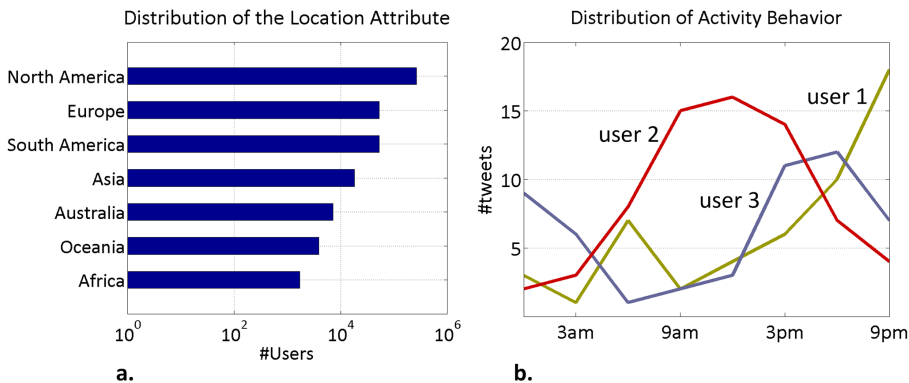

Figure 3: (a) Distribution of Twitter users in different continents based on the location attribute; (b) Comparative examples of activity behavior distributions of three users over a period of 24 hours.

**Location of users**, extracted using the timezone attribute of Twitter users. Each timezone was mapped to the continent it belonged to (e.g. Asia, Europe and North America (ref. Figure 3(a)); so that we had a coarse sense of the location of the users. Note that although Twitter provides a field for the location of the users, we rather decided to use the timezone information, as we found it to be less noisy and less sparse.

**Information roles of users**, we consider three categories of roles: "generators", "mediators" and "receptors" (Figure 4). Generators are users who create several posts (or tweets) but few users respond to them (via the @ tag on Twitter, which is typically used with the username to respond to a particular user, e.g. @BillGates). While receptors are those who create fewer posts but receive several posts as responses. Mediators are users who lie between these two categories.

---
[2]For simplicity, we omit specifying the attribute value $v$ in the rest of the article, and refer to $G(a_k = v)$ as the "attribute social graph" $G(a_k)$.

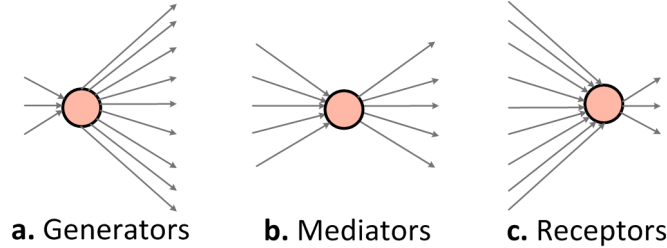

**a.** Generators   **b.** Mediators   **c.** Receptors

**Figure 4:** **The three categories of user behavior corresponding to the attribute: information roles. Out-going links indicate post creation while incoming links indicate post responses from other users.**

**Content creation of users**, we use the two content creation roles: "meformer" (users who primarily post content relating to self) and "informer" (users posting content about external happenings) as discussed in [24].

**Activity behavior of users**, i.e. the distribution of a particular social action over a certain time period. We consider the mean number of posts (tweets) per user over 24 hours (ref. Figure 3(b)) and compute similarities between pairs of users based on the Kullback-Leibler (KL) divergence measure of comparing across distributions.

### 3.1.2 Attribute Homophily

Attribute homophily [23, 22] is defined as the tendency of users in a social graph to associate and bond with others who are "similar" to them along a certain attribute or contextual dimension e.g. age, gender, race, political view or organizational affiliation. Specifically, a pair of users can be said to be "homophilous" if one of their attributes match in a proportion greater than that in the network of which they are a part. Hence in our context, for a particular value of $a_k \in \mathcal{A}$, the users in the social graph $G(a_k)$ corresponding to that value are homophilous to each other.

### 3.1.3 Topic Diffusion

Diffusion with respect to a particular topic at a certain time is given as the flow of information on the topic from one user to another via the social graph, and based on a particular social action. Specifically,

**Definition 1** *Given two users $u_i$ and $u_j$ in the baseline social graph $G$ such that $e_{ij} \in E$, there is diffusion of information on topic $\theta$ from $u_j$ to $u_i$ if $u_j$ performs a particular social action $O_r$ related to $\theta$ at a time slice $t_{m-1}$ and is* *succeeded by $u_i$ in performing the same action on $\theta$ at the next time slice $t_m$, where $t_{m-1} < t_m$.*[3]

Further, topic diffusion subject to homophily along the attribute $a_k$ is defined as the diffusion over the attribute social graph $G(a_k)$.

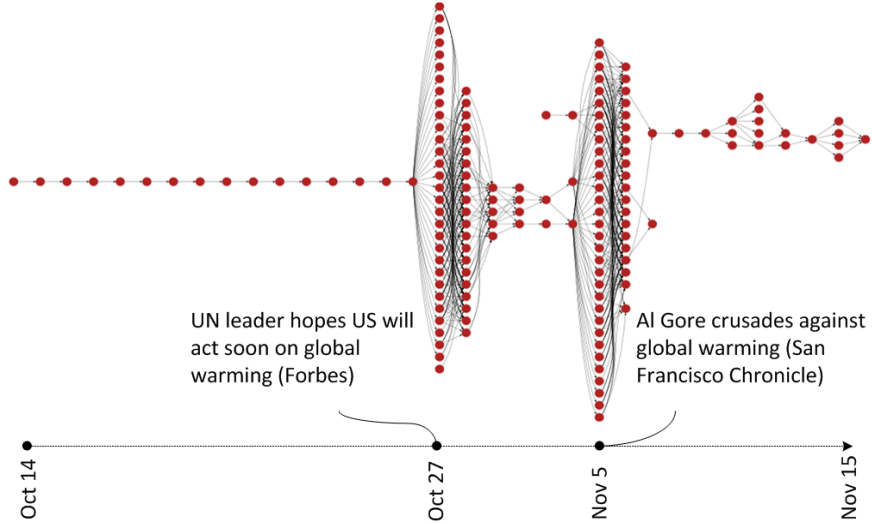

Figure 5: Example of a diffusion series from Twitter on the topic "global warming". The nodes are users involved in diffusion while the edges represent "friend links" connecting two users.

In the context of Twitter, topic diffusion can manifest itself through three types of evidences: (1) users posting tweets using the same URL, (2) users tweeting with the same hashtag (e.g. #MichaelJackson) or a set of common keywords, and (3) users using the re-tweet (RT) symbol. We utilize all these three cases of topic diffusion in this work. Also note, the hashtag or the common set of keywords associated with the tweets are assumed to be the topics in this article.

### 3.1.4 Diffusion Series

In order to characterize diffusion, we now define a topology called a *diffusion series*[4] that summarizes diffusion in a social graph for a given topic over a period of time [10]. Formally,

**Definition 2** *A diffusion series $s_N(\theta)$ on topic $\theta$ and over time slices $t_1$ to $t_N$ is defined as a directed acyclic graph where the nodes represent a subset of users in the baseline social graph $G$, who are involved in a specific social action $O_r$ over $\theta$ at any time slice between $t_1$ and $t_N$.*

---

[3]Since we discuss our problem formulation and methodology for a specific social action, the dependence of different concepts on $O_r$ is omitted in the rest of the article for simplicity.

[4]Note, a diffusion series is similar to a diffusion tree as in [21, 3], however we call it a "series" since it is constructed progressively over a period of time and allows a node to have multiple sources of diffusion.

Note, in a diffusion series $s_N(\theta)$ a node represents an occurrence of a user $u_i$ creating at least one instance of the social action $O_r$ about $\theta$ at a certain time slice $t_m$ such that $t_1 \leq t_m \leq t_N$. Nodes are organized into "slots"; where nodes associated with the same time slice $t_m$ are arranged into the same slot $l_m$. Hence it is possible that the same user is present at multiple slots in the series if s/he tweets about the same topic $\theta$ at different time slices. Additionally, there are edges between nodes across two adjacent slots, indicating that user $u_i$ in slot $l_m$ performs the social action $O_r$ on $\theta$ at $t_m$, after her friend $u_j$ has performed action on the same topic $\theta$ at the previous time slice $t_{m-1}$ (i.e. at slot $l_{m-1}$). There are no edges between nodes at the same slot $l_m$: a diffusion series $s_N(\theta)$ in this work captures diffusion on topic $\theta$ *across* time slices, and does not include possible flow occurring at the same time slice.

For the Twitter dataset, we have chosen the granularity of the time slice $t_m$ to be sufficiently small, i.e. a day to capture the dynamics of diffusion. Thus all the users at slot $l_m$ tweet about $\theta$ on the same day; and two consecutive slots have a time difference of one day. An example of a diffusion series on Twitter over the topic "global warming" has been shown in Figure 5, qualitatively annotated by the authors with significant relevant news events (http://www.news.google.com/). Several other examples of diffusion series over a set of diverse "trending topics" from Twitter are also shown in Figure 6. Note that the topology of each series is markedly different for each case.

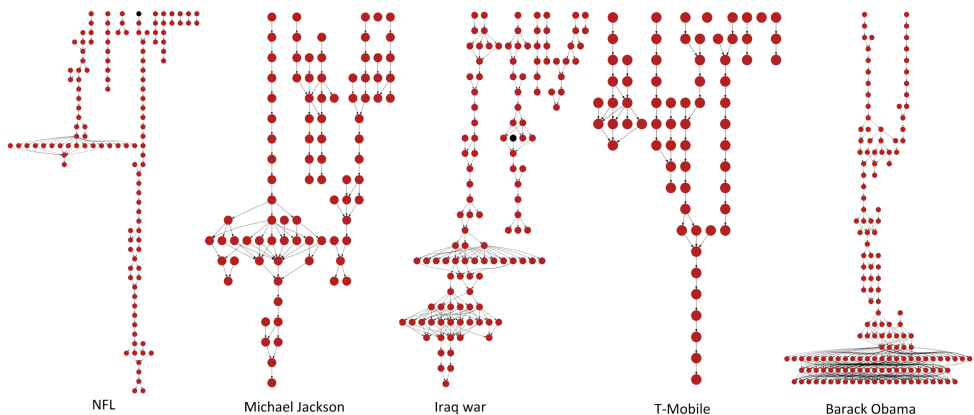

Figure 6: Example of diffusion series for several "trending topics" from Twitter during the time period between Sep-Nov 2009.

Since each topic $\theta$ can have multiple disconnected diffusion series $s_N(\theta)$ at any given time slice $t_N$, we call the family of all diffusion series a *diffusion collection* $\mathcal{S}_N(\theta) = \{s_N(\theta)\}$. Corresponding to each value of the attribute $a_k$, the diffusion collection over the attribute social graph $G(a_k)$ at $t_N$ is similarly given as $\mathcal{S}_{N;a_k}(\theta) = \{s_{N;a_k}(\theta)\}$.

## 3.2 Problem Statement

Given, (1) a baseline social graph $G(V, E)$; (2) a set of social actions $\mathcal{O} = \{O_1, O_2, \ldots\}$ that can be performed by users in $V$, and (3) a set of attributes $\mathcal{A} = \{a_k\}$ that are shared by users in $V$, we perform the following two preliminary steps. First, we construct the attribute social graphs $\{G(a_k)\}$, for all values of $a_k \in \mathcal{A}$. Second, we construct diffusion collections corresponding to $G$ and $\{G(a_k)\}$ for a given topic $\theta$ (on which diffusion is to be estimated over time slices $t_1$ to $t_N$) and a particular social action $O_r$: these are given as $\mathcal{S}_N(\theta)$ and $\{\mathcal{S}_{N;a_k}(\theta)\}$ respectively. The technical problem addressed in this article involves the following:

1. *Characterization:* Based on each of the diffusion collections $\mathcal{S}_N(\theta)$ and $\{\mathcal{S}_{N;a_k}(\theta)\}$, we extract diffusion characteristics on $\theta$ at time slice $t_N$ given as: $\mathbf{d}_N(\theta)$ and $\{\mathbf{d}_{N;a_k}(\theta)\}$ respectively (section 4);

2. *Prediction:* We predict the set of users likely to perform the same social action at the next time slice $t_{N+1}$ corresponding to each of the diffusion collections $\mathcal{S}_N(\theta)$ and $\{\mathcal{S}_{N;a_k}(\theta)\}$. This gives the diffusion collections at $t_{N+1}$: $\hat{\mathcal{S}}_{N+1}(\theta)$ and $\{\hat{\mathcal{S}}_{N+1;a_k}(\theta)\} \forall a_k \in \mathcal{A}$ (section 5);

3. *Distortion Measurement:* We extract diffusion characteristics at $t_{N+1}$ over the (predicted) diffusion collections, $\hat{\mathcal{S}}_{N+1}(\theta)$ and $\{\hat{\mathcal{S}}_{N+1;a_k}(\theta)\}$, given as, $\hat{\mathbf{d}}_{N+1}(\theta)$ and $\{\hat{\mathbf{d}}_{N+1;a_k}(\theta)\}$ respectively. Now we quantify the impact of attribute homophily on diffusion based on two kinds of distortion measurements on $\hat{\mathbf{d}}_{N+1}(\theta)$ and $\{\hat{\mathbf{d}}_{N+1;a_k}(\theta)\}$. A particular attribute $a_k \in \mathcal{A}$ would have an impact on diffusion if $\hat{\mathbf{d}}_{N+1;a_k}(\theta)$, avergaed over all possible values of $a_k$: (a) has lower distortion with respect to the actual (i.e. $\mathbf{d}_{N+1}(\theta)$); and (b) can quantify external time series (search, news trends) better, compared to either $\hat{\mathbf{d}}_{N+1}(\theta)$ or $\{\hat{\mathbf{d}}_{N+1;a'_k}(\theta)\}$, where $k' \neq k$ (section 6).

## 4 Characterizing Diffusion

We describe eight different measures for quantifying diffusion characteristics given by the baseline and the attribute social graphs on a certain topic and via a particular social action [10]. The measures are categorized through various aspects such as: properties of users involved in diffusion (volume, participation and dissemination), diffusion series topology (reach, spread, cascade instances and collection size) and temporal properties (rate). We discuss the measures for the diffusion collection corresponding to the baseline social graph (i.e. $\mathcal{S}_N(\theta)$); the computation of these measures on the attribute social graphs follow correspondingly over their respective diffusion collections (i.e. $\{\mathcal{S}_{N;a_k}(\theta)\}$).

**Volume**: Volume is a notion of the overall degree of contagion in the social graph. For the diffusion collection $\mathcal{S}_N(\theta)$ over the baseline social graph $G$, we formally define volume $v_N(\theta)$ with respect to $\theta$ and at time slice $t_N$ as the ratio of $n_N(\theta)$ to $\eta_N(\theta)$, where $n_N(\theta)$ is the total number of users (nodes) in the diffusion collection $\mathcal{S}_N(\theta)$, and $\eta_N(\theta)$ is the number of users in the social graph $G$ associated with $\theta$. Note, $n_N(\theta)$ would include users who are not part of the diffusion collection, but nevertheless have tweeted about $\theta$.

**Participation**: Participation $p_N(\theta)$ at time slice $t_N$ [3] is the fraction of users involved in the diffusion of information on a particular topic who further trigger other users in the social graph to get involved in the diffusion. It is the ratio of the number of non-leaf nodes in the diffusion collection $\mathcal{S}_N(\theta)$, normalized by

$\eta_N(\theta)$.

**Dissemination**: Dissemination $\delta_N(\theta)$ at time slice $t_N$ is given by the ratio of the number of users in the diffusion collection $\mathcal{S}_N(\theta)$ who do not have a parent node, normalized by $\eta_N(\theta)$. In other words, they are the "seed users" or ones who get involved in the diffusion due to some unobservable external influence, e.g. a news event.

**Reach**: Reach $r_N(\theta)$ at time slice $t_N$ [21] is conceptually defined as the extent in the social graph, to which information on a particular topic $\theta$ reaches to users. We define it formally as the ratio of the mean of the number of slots to the sum of the number of slots in all diffusion series belonging to $\mathcal{S}_N(\theta)$.

**Spread**: For the diffusion collection $\mathcal{S}_N(\theta)$, spread $s_N(\theta)$ at time slice $t_N$ [21] is defined as the ratio of the maximum number of nodes at any slot in $s_N(\theta) \in \mathcal{S}_N(\theta)$ to $n_N(\theta)$.

**Cascade Instances**: Cascade instances $c_N(\theta)$ at time slice $t_N$ is defined as the ratio of the number of slots in the diffusion series $s_N(\theta) \in \mathcal{S}_N(\theta)$ where the number of *new* users at a slot $l_m$ (i.e. non-occurring at a previous slot) is greater than that at the previous slot $l_{m-1}$, to $L_N(\theta)$, the number of slots in $s_N(\theta) \in \mathcal{S}_N(\theta)$.

**Collection Size**: Collection size $\alpha_N(\theta)$ at time slice $t_N$ is the ratio of the number of diffusion series $s_N(\theta)$ in $\mathcal{S}_N(\theta)$ over topic $\theta$, to the total number of connected components in the social graph $G$.

**Rate**: We define rate $\gamma_N(\theta)$ at time slice $t_N$ as the "speed" at which information on $\theta$ diffuses in the collection $\mathcal{S}_N(\theta)$. It depends on the difference between the median time of posting of tweets at all consecutive slots $l_m$ and $l_{m-1}$ in the diffusion series $s_N(\theta) \in \mathcal{S}_N(\theta)$. Hence it is given as:

$$\gamma_N(\theta) = 1/(1 + \frac{1}{L_N(\theta)} \sum_{l_{m-1}, l_m \in \mathcal{S}_N(\theta)} (\bar{t}_m(\theta) - \bar{t}_{m-1}(\theta)), \qquad (1)$$

where $\bar{t}_m(\theta)$ and $\bar{t}_{m-1}(\theta)$ are measured in seconds and $\bar{t}_m(\theta)$ corresponds to the median time of tweet at slot $l_m$ in $s_N(\theta) \in \mathcal{S}_N(\theta)$.

These diffusion measures thus characterize diffusion at time slice $t_N$ over $\mathcal{S}_N(\theta)$ as the vector: $\mathbf{d}_N(\theta) = [v_N(\theta), p_N(\theta), \delta_N(\theta), r_N(\theta), s_N(\theta), c_N(\theta), \alpha_N(\theta), \gamma_N(\theta)]$. Similarly, we compute the diffusion measures vector over $\{\mathcal{S}_{N;a_k}(\theta)\}$, given by: $\{\mathbf{d}_{N;a_k}(\theta)\} = \{[v_{N;a_k}(\theta), p_{N;a_k}(\theta), \delta_{N;a_k}(\theta), r_{N;a_k}(\theta), s_{N;a_k}(\theta), c_{N;a_k}(\theta), \alpha_{N;a_k}(\theta), \gamma_{N;a_k}(\theta)]\}$, corresponding to each value of $a_k$.

## 5 Prediction Framework

In this section we present our method of predicting the users who would be part of the diffusion collections at a future time slice for the baseline and attribute social graphs. Our method comprises the following steps. (1) Given the observed diffusion collections until time slice $t_N$ (i.e. $\mathcal{S}_N(\theta)$ and $\mathcal{S}_{N;a_k}(\theta)$), we first propose a probabilistic framework based on Dynamic Bayesian networks [25] to predict the users likely to perform the social action $O_r$ at the next time slice $t_{N+1}$. This would yield us users at slot $l_{N+1}$ in the different diffusion series at $t_{N+1}$. (2) Next, these predicted users give the diffusion collections at $t_{N+1}$: $\hat{\mathcal{S}}_{N+1}(\theta)$ and $\{\hat{\mathcal{S}}_{N+1;a_k}(\theta)\}$.

We present a Dynamic Bayesian network (DBN) representation of a particular social action by a user over time, that helps us predict the set of users likely to perform the social action at a future time (Figure 7(a)). Specifically, at any time slice $t_N$, a given topic $\theta$ and a given social action, the DBN captures the relationship between three nodes:

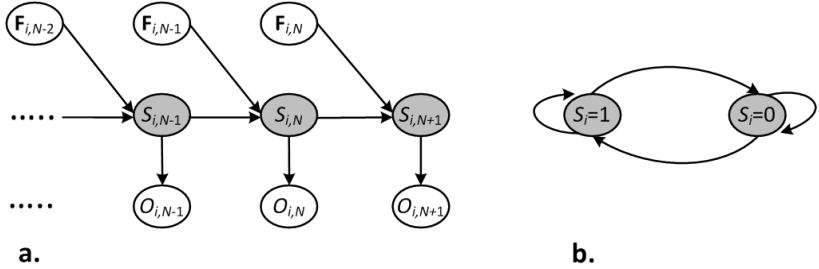

Figure 7: (a) Structure of the Dynamic Bayesian network used for modeling social action of a user $u_i$. The diagram shows the relationship between environmental features ($\mathbf{F}_{i,N}(\theta)$), hidden states ($S_{i,N}(\theta)$) and the observed action ($O_{i,N}(\theta)$). (b) State transition diagram showing the 'vulnerable' ($S_i = 1$) and 'indifferent' states ($S_i = 0$) of a user $u_i$.

**Environmental Features.** That is, the set of contextual variables that effect a user $u_i$'s decision to perform the action on $\theta$ at a future time slice $t_{N+1}$ (given by $\mathbf{F}_{i,N}(\theta)$). It comprises three different measures: (1) $u_i$'s degree of activity on $\theta$ in the past, given as the ratio of the number of posts (or tweets) by $u_i$ on $\theta$, to the total number of posts between $t_1$ and $t_N$; (2) mean degree of activity of $u_i$'s friends in the past, given as the ratio of the number of posts by $u_i$'s friends on $\theta$, to the total number of posts by them between $t_1$ and $t_N$; and (3) popularity of topic $\theta$ at the previous time slice $t_N$, given as the ratio of the number of posts by all users on $\theta$, to the total number of posts at $t_N$.

**States.** That is, latent states ($S_{i,N}(\theta)$) of the user $u_i$ responsible for her involvement in diffusion at $t_{N+1}$. Our motivation in conceiving the latent states comes from the observation that, in the context of Twitter, a user can tweet on a topic under two kinds of circumstances: first, when she observes her friend doing so already: making her *vulnerable* to diffusion; and second, when her tweeting is *indifferent* to the activities of her friends. Hence the state node at $t_{N+1}$ that impacts $u_i$'s action can have two values as the vulnerable and the indifferent state (Figure 7(b)).

**Observed Action.** That is, evidence ($O_{i,N}(\theta)$) of the user $u_i$ performing (or not performing) the action, corresponding values being: $\{1, 0\}$ respectively. It is the observable output of hidden state node $S_{i,N+1}(\theta)$ and is affected by the environmental features $\mathbf{F}_{i,N}(\theta)$ at the previous time slice $t_N$, assuming first order Markov property in the DBN.

Now we show how to predict the probability of the observed action at $t_{N+1}$ (i.e. $\hat{O}_{i,N+1}(\theta)$) using $\mathbf{F}_{i,N}(\theta)$ and $S_{i,N+1}(\theta)$, based on the DBN model. Our goal is to estimate the following expectation[5]:

$$\hat{O}_{i,N+1} = E(O_{i,N+1}|O_{i,N}, \mathbf{F}_{i,N}). \tag{2}$$

This involves computing $P(O_{i,N+1}|O_{i,N}, \mathbf{F}_{i,N})$. This conditional probability can be written as an inference equation using the temporal dependencies given by the DBN and assuming first order Markov property:

$$\begin{aligned}
&P(O_{i,N+1}|O_{i,N}, \mathbf{F}_{i,N}) \\
&= \sum_{S_{i,N+1}} \left[ P(O_{i,N+1}|S_{i,N+1}, O_{i,N}, \mathbf{F}_{i,N}).P(S_{i,N+1}|O_{i,N}, \mathbf{F}_{i,N}) \right]. \\
&= \sum_{S_{i,N+1}} P(O_{i,N+1}|S_{i,N+1}).P(S_{i,N+1}|S_{i,N}, \mathbf{F}_{i,N}).
\end{aligned} \tag{3}$$

Our prediction task thus involves two parts: predicting the probability of the hidden states given the environmental features, $P(S_{i,N+1}|S_{i,N}, \mathbf{F}_{i,N})$; and predicting the probability density of the observation nodes given the hidden states, $P(O_{i,N+1}|S_{i,N+1})$, and thereby the expected value of observation nodes $\hat{O}_{i,N+1}$. These two steps are discussed in the following subsections.

## 5.1 Predicting Hidden States

Using Bayes rule, we apply conditional independence between the hidden states and the environmental features at the same time slice (ref. Figure 7(a)). The probability of the hidden states at $t_{N+1}$ given the environmental features at $t_N$, i.e. $P(S_{i,N+1}|S_{i,N}, \mathbf{F}_{i,N})$ can be written as:

$$P(S_{i,N+1}|S_{i,N}, \mathbf{F}_{i,N}) \propto P(\mathbf{F}_{i,N}|S_{i,N}).P(S_{i,N+1}|S_{i,N}). \tag{4}$$

Now, to estimate the probability density of $P(S_{i,N+1}|S_{i,N}, \mathbf{F}_{i,N})$ using eqn. 4 we assume that the hidden states $S_{i,N+1}$ follows a multinomial distribution over the environmental features $\mathbf{F}_{i,N}$ with parameter $\phi_{i,N}$, and a conjugate Dirichlet prior over the previous state $S_{i,N}$ with parameter $\lambda_{i,N+1}$. The optimal parameters of the pdf of $P(S_{i,N+1}|S_{i,N}, \mathbf{F}_{i,N})$ can now be estimated using MAP:

---

[5]Without loss of generalization, we omit the topic $\theta$ in the variables in this subsection for the sake of simplicity.

$$\begin{aligned}
&\mathcal{L}(P(S_{i,N+1}|S_{i,N}, \mathbf{F}_{i,N})) \\
&= \log(P(\mathbf{F}_{i,N}|S_{i,N})) + \log(P(S_{i,N+1}|S_{i,N})) \\
&= \log \mathbf{multinom}(\text{vec}(\mathbf{F}_{i,N}); \phi_{i,N}) \\
&\quad + \log \mathbf{Dirichlet}(\text{vec}(S_{i,N+1}); \lambda_{i,N+1}) \\
&= \log \frac{\sum_{jk} \mathbf{F}_{i,N;jk}!}{\prod_{jk} \mathbf{F}_{i,N;jk}!} \prod_{jk} \phi_{i,N;jk}^{\mathbf{F}_{i,N;jk}} + \log \frac{1}{B(\lambda_{i,N+1})} \prod_{jl} S_{i,N+1}^{S_{i,N;jl}} \\
&= \sum_{jk} \mathbf{F}_{i,N;jk} . \log \phi_{i,N;jk} + \sum_{jl} S_{i,N;jl} . \log S_{i,N+1;jl} + \text{const.}
\end{aligned} \quad (5)$$

where $B(\lambda_{i,N+1})$ is a beta-function with the parameter $\lambda_{i,N+1}$. Maximizing the log likelihood in eqn 5 hence yields the optimal parameters for the pdf of $P(S_{i,N+1}|S_{i,N}, \mathbf{F}_{i,N})$. The details of the convergence of the above estimation have been skipped and can be found in [25].

### 5.2 Predicting Observed Action

To estimate the probability density of the observation nodes given the hidden states, i.e. $P(O_{i,N+1}|S_{i,N+1})$ we adopt a generative model approach and train two discriminative Hidden Markov Models—one corresponding to the class when $u_i$ performs the action, and the other when she does not. Based on observed actions from $t_1$ to $t_N$, we learn the parameters of the HMMs using the Baum-Welch algorithm. We then use the emission probability $P(O_{i,N+1}|S_{i,N+1})$ given by the observation-state transition matrix to determine the most likely sequence at $t_{N+1}$ using the Viterbi algorithm. The details of this estimation can be found in [27]. We finally substitute the emission probability $P(O_{i,N+1}|S_{i,N+1})$ from above and $P(S_{i,N+1}|S_{i,N}, \mathbf{F}_{i,N})$ from eqn. 5 into eqn. **??** to compute the expectation $E(O_{i,N+1}|O_{i,N}, \mathbf{F}_{i,N})$ and get the estimated observed action of $u_i$: $\hat{O}_{i,N+1}$ (eqn. 2).

We now use the estimated social actions $\hat{O}_{i,N+1}(\theta)$ of all users at time slice $t_{N+1}$ to get a set of users who are likely to involve in the diffusion process at $t_{N+1}$ for both the baseline and the attribute social graphs. Next we use $G$ and $\{G(a_k)\}$ to associate edges between the predicted user set, and the users in each diffusion series corresponding to the diffusion collections at $t_N$. This gives the diffusion collection $t_{N+1}$, i.e. $\hat{\mathcal{S}}_{N+1}(\theta)$ and $\{\hat{\mathcal{S}}_{N+1;a_k}(\theta)\}$ (ref. section 3.1.4).

## 6 Distortion Measurement

We now compute the diffusion feature vectors $\hat{\mathbf{d}}_{N+1}(\theta)$ or $\{\hat{\mathbf{d}}_{N+1;a_k}(\theta)\}$ based on the predicted diffusion collections $\hat{\mathcal{S}}_{N+1}(\theta)$ and $\{\hat{\mathcal{S}}_{N+1;a_k}(\theta)\}$ from section 5. To quantify the impact of attribute homophily on diffusion at $t_{N+1}$ corresponding to $a_k \in \mathcal{A}$, we define two kinds of distortion measures—(1) saturation measurement, and (2) utility measurement metrics.

**Saturation Measurement.** We compare distortion between the predicted and actual diffusion characteristics at $t_{N+1}$. The saturation measurement metric is thus given as $1 - D(\hat{\mathbf{d}}_{N+1}(\theta), \mathbf{d}_{N+1}(\theta))$ and $1 - D(\hat{\mathbf{d}}_{N+1;a_k}(\theta), \mathbf{d}_{N+1}(\theta))$, averaged over all values of $\forall a_k \in \mathcal{A}$ respectively for the baseline and the attribute social graphs. $\mathbf{d}_{N+1}(\theta)$ gives the actual diffusion characteristics at $t_{N+1}$ and $D(A, B)$ Kolmogorov-Smirnov (KS) statistic, defined as $max(|A - B|)$.

**Utility Measurement.** Note that in social media, both the search for information about events and important news events are typically empirically related to information diffusion. That is, our intuition is that diffusion is causally related to external phenomena, such as search behavior or appearance of news articles on a certain topic. A valid model of diffusion should therefore not only be able to predict diffusion, but the predicted diffusion also ought to be able to exhibit correlation to the external phenomena such as information search and world news trends. If the predicted diffusion on an attribute shows high correlation with external trends, it implies its utility in the discovery of diffusion.

Hence we describe two utility measurement metrics for quantifying the relationship between the predicted diffusion characteristics $\hat{\mathbf{d}}_{N+1}(\theta)$ or $\{\hat{\mathbf{d}}_{N+1;a_k}(\theta)\}$ on topic $\theta$, and the trends of same topic $\theta$ obtained from external time series [10]. We collect two kinds of external trends: (1) *search trends*–the search volume of the topic $\theta$ over a period of time from $t_1$ to $t_{N+1}$[6]; (2) *news trends*—the frequency of archived news articles about the same topic $\theta$ over same period[7]. The utility measurement metrics are defined as follows:

1. *Search trend measurement*: We first compute the cumulative distribution function (CDF) of diffusion volume as $E_{N+1}^D(\theta) = \sum_{m \leq (N+1)} |l_m(\hat{\mathcal{S}}_{N+1}(\theta))|/Q_D$, where $|l_m(\hat{\mathcal{S}}_{N+1}(\theta))|$ is the number of nodes at slot $l_m$ in the collection $\hat{\mathcal{S}}_{N+1}(\theta)$. $Q_D$ is the normalized term and is defined as $\sum_m |l_m(\hat{\mathcal{S}}_{N+1}(\theta))|$. Next, we compute the CDF of search volume as $E_{N+1}^S(\theta) = \sum_{m \leq (N+1)} f_m^S(\theta)/Q_S$, where $f_m^S(\theta)$ is the search volume at $t_m$, and $Q_S$ is the normalization term. The search trend measurement is defined as $1 - D(E_{N+1}^D(\theta), E_{N+1}^S(\theta))$, where $D(A, B)$ is the KS statistic.

2. *News trend measurement*: Similarly, we compute the CDF of news volume as $E_{N+1}^{\mathcal{N}}(\theta) = \sum_{m \leq (N+1)} f_m^{\mathcal{N}}(\theta)/Q_{\mathcal{N}}$, where $f_m^{\mathcal{N}}(\theta)$ is the number of archived news articles available from Google News for $t_m$, and $Q_{\mathcal{N}}$ is the normalization term. The news trend measurement is similarly defined as $1 - D(E_{N+1}^D(\theta), E_{N+1}^{\mathcal{N}}(\theta))$.

Using the same method as above, we compute the search and news trend measurement metrics for the attribute social graphs—given as, $1 - D(E_{N+1;a_k}^D(\theta), E_{N+1}^S(\theta))$ and $1 - D(E_{N+1;a_k}^D(\theta), E_{N+1}^{\mathcal{N}}(\theta))$, averaged over all values of $\forall a_k \in \mathcal{A}$ respectively.

## 7  Experimental Results

We present our experimental results in this section. First we discuss data preparation (section 7.1), followed by analysis of saturation and utility measurement of diffusion characteristics at the quantitative level: time-based, theme-based and comparative evaluation against baseline methods (section 7.2). We finally discuss some of the implications of this work in section 7.3.

---

[6] http://www.google.com/intl/en/trends/about.html
[7] http://news.google.com/

Table 1: **Summary of statistics of the data used for studying diffusion on Twitter.**

| ATTRIBUTE | VALUE |
|---|---|
| #nodes | 465,107 |
| #edges | 836,541 |
| #nodes with time-zone attribute | 385,547 |
| #tweets | 25,378,846 |
| Time span of tweets' post times | Oct'06–Nov'09 |

Table 2: **Example theme and trending topic associations on Twitter dataset.**

| THEME | TRENDING TOPICS |
|---|---|
| **Politics** | Obama, Senate, Afghanistan, Tehran, Healthcare |
| **Entertainment-Culture** | Beyonce, Eagles, Michael Jackson, #britney3premiere |
| **Sports** | Chargers, Cliff Lee, Dodgers, Formula One, New York Yankees |
| **Technology-Internet** | Android 2, Bing, Google Wave, Windows 7, #Firefox5 |
| **Social Issues** | Swine Flu, Unemployment, #BeatCancer, #Stoptheviolence |

## 7.1 Data Preparation

**Twitter Dataset.** We have focused on a large dataset crawled from Twitter. We have undertaken a focused crawl[8] based on a snowballing technique, over a set of quality users (∼465K), who mutually form a reasonably large connected component. First, we seeded the crawl from a set of genuine (or authoritative) users, who post about a diverse range of topics and reasonably frequently. Our seed set size is 500; and comprises politicians, musicians, environmentalists, techies and so on. These lists were collected from the popular social media blog, Mashable (http://mashable.com/2008/10/20/25-celebrity-twitter-users/). Next we expand the social graph from the seed set based on their "friend" links[9]. We finally executed a dedicated cron job that collected the tweets (and their associated timestamps) for users in the entire social graph every 24 hours. Table I gives some basic statistics of the crawled data spanning over a three year period[10].

---

[8]http://apiwiki.twitter.com/
[9]Note that the social graph crawled in this work is a static snapshot made only once at the time of the crawl.
[10]The dataset is available for download at http://www.public.asu.edu/ mdechoud/datasets.html

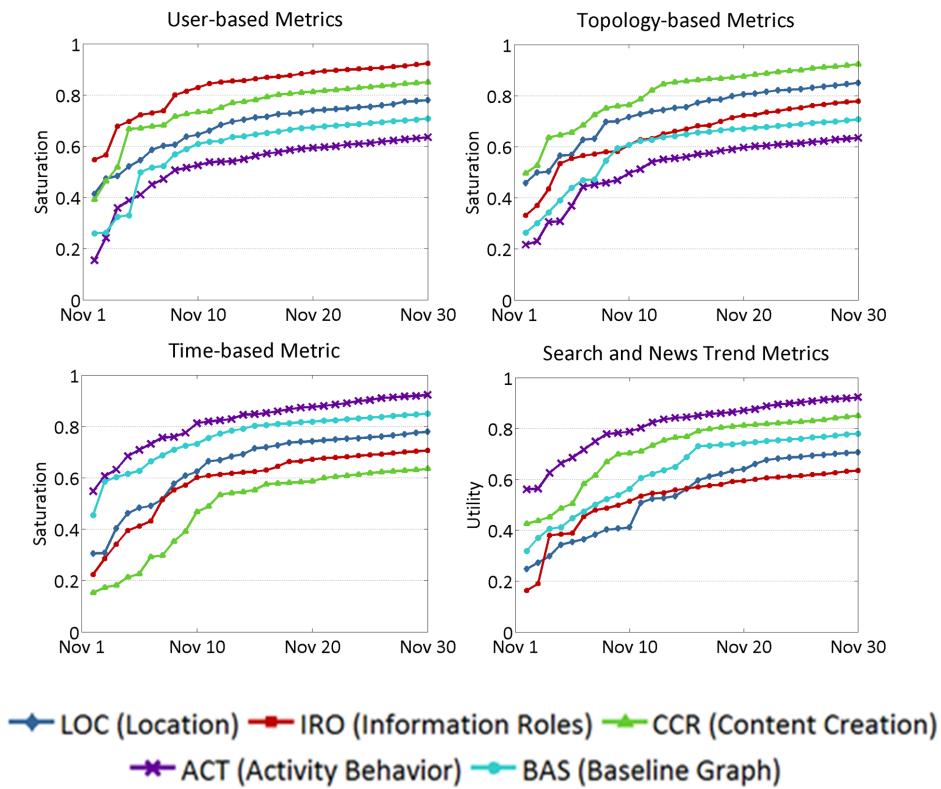

Figure 8: Performance analysis of saturation and utility measurement of predicted diffusion characteristics over time (the higher, the better). Results are shown across different categories of saturation measurement metrics: user-based (volume, participation, dissemination), topology-based (reach, spread, cascade instances, collection size), time-based (rate), utility measurement metrics: search and news trends.

**Experimental Setup.** The crawled social graph, comprising the users and their tweets are now deployed in the study of diffusion. Since we are interested in studying diffusion at the granularity of a topic, we first define how we conceive of the topics. For our experiments, we focus on the "trending topics"[11] that are featured on Twitter over a two month period between Sep and Nov 2009. From the ensemble of these trending topics, a set of $\sim 125$ topics are selected at random; of which there are 25 hashtags and the rest, phrases or groups of words.

For the ease of analysis, we organize the different trending topics into generalized themes. For automatically assigning theme to trending topic associations, we use the popular open source natural language processing toolkit called "OpenCalais"[12] e.g. Table II. In the context of Twitter, we filter tweets give a trending topic, and then use OpenCalais to return theme labels over those tweets. Based on this process, we associated the 125 trending topics with a total of nine themes, such as 'Business Finance', 'Sports' etc'.

Now our experimental goal is to utilize the crawled social graph to construct the baseline and the attribute social graphs, predict diffusion characteristics, and then study these characteristics over time, subject to homophily on each of their respective diffusion collections. For the purpose, we adopt a "batch" method of incremental training and testing. We begin with a base training set size comprising tweets posted during Sep 2009, and then incrementally train and predict the diffusion characteristics over Oct and Nov 2009. In the subsequent subsections, the predicted diffusion characteristics are discussed over the said period.

## 7.2 Quantitative Analysis

Now we present quantitative analysis of impact of attribute homophily on diffusion. First we present the variations of predicted diffusion characteristics corresponding to different attributes, over diffusion metrics and time. Second, we discuss how prediction performance over different attributes varies across different themes. Finally we present a comparative study of the performance of our proposed method against several baseline techniques.

### 7.2.1 Temporal Analysis across Diffusion Metrics

Figure 8 presents the temporal variations of the performance of predicted diffusion characteristics based on saturation and utility measurements, averaged over all eight themes. We organize the results corresponding to different categories of metrics: user-based (volume, participation, dissemination), topology-based (reach, spread, cascade instances, collection size), time-based (rate), and external time-series variables such as search and news trends.

The observations from the results reveal interesting insights. Overall, firstly, as is intuitive, we observe that as we increase the training data size (i.e. over time), the saturation and utility measure increase for the case of all attributes. The results in Figure 8 indicate that compared to the predictions over the baseline social graphs, several attribute social graphs yield higher saturation and utility measures in explaining the actual diffusion characteristics as well as the external trends respectively. However, note that the attributes corresponding to the best and the worst performance vary across the metrics:

- **User-based metrics.** For the user-based metrics, we observe that the attribute, information roles (IRO) yields the highest saturation over time. This is because diffusion characteristics such as volume and participation are often related to the information generation and consumption behavior of users, and hence the higher performance. The worst performance in this case corresponds to the activity behavior attribute (ACT), revealing that the user-based metrics are less affected by the time of the day corresponding to the tweeting activity.

- **Topology-based metrics.** In the case of the topology-based metrics, the attribute, content creation (CCR) yields the best performance over the month of prediction. This is because diffusion characteristics such as reach and spread are often affected by how much the information in the tweets are associated with external happenings. Since the content creation attribute characterizes users based on the type of content they share via tweets, it explains the high saturation measures observed in this case. Note, the location attribute (LOC) also yields good performance. This is because the topology metrics are often related to the social connectivity in the social graph; and location is likely to play a significant role in defining the social connectivities in the social graph under consideration.

- **Time-based metric.** The attribute activity behavior (ACT) corresponds to the highest saturation measures in the case of the time-based metric rate. This is because the "speed" of information diffusing in the network often depends upon the temporal pattern of activity of the users, i.e. when are they tweeting. Interestingly, in this case the baseline social graph (BAS) corresponds to a better performance compared to several other attributes. We conjecture that it reveals that rate is less affected by our chosen set of user attributes.

- **Search and news trends.** Finally, in the case of search and news trends, the best performance again corresponds to the activity behavior attribute (ACT); implying that search behavior and response to news events are often affected highly by the temporal patterns of posting of tweets by users. Moreover, LOC attribute's performance is relatively poor in this case; for which we conjecture that search or news trends are usually less affected by the location of the users.

---
[11]Trending topics are Twitter-generated list of popular topics. Note they can either be hashtags (i.e. words or phrases preceded by the # symbol), or could be groups of words.
[12]http://www.opencalais.com/

To summarize, our primary insight from these experiments is that the particular attribute thats yields the best prediction over the diffusion process often depends upon the metric under consideration.

### 7.2.2 Analysis across Themes

Now we discuss attribute homophily subject to variations across the different themes, and averaged over time (Oct-Nov 2009). Figure 9 shows that there is considerable variation in performance (in terms of saturation and utility measures) over the eight themes.

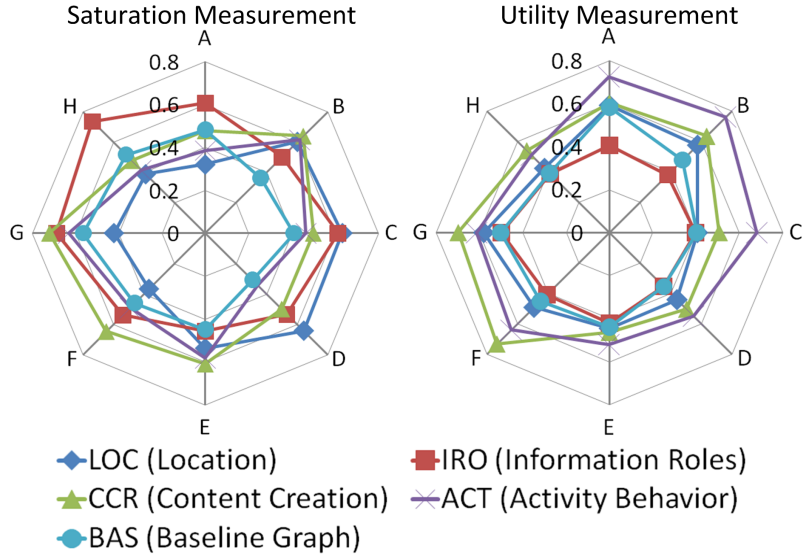

Figure 9: Mean saturation and utility measurement of predicted diffusion characteristics shown across different themes. The themes are: A–Business-Finance, B–Politics, C–Entertainment-Culture, D–Sports, E–Technology-Internet, F–Human Interest, G–Social Issues, H–Hospitality-Recreation.

In the case of saturation measurement, we observe that the location attribute (LOC) yields high saturation measures over themes related to events that are often "local" in nature: e.g. (1) 'Sports' comprising topics such as 'NBA', 'New York Yankees', 'Chargers', 'Sehwag' and so on–each of them being of interest to users respectively from the US, NYC, San Diego and India; and (2) 'Politics' (that includes topics like 'Obama', 'Tehran' and 'Afghanistan')—all of which were associated with important, essentially local happenings during the period of our analysis. Whereas for themes that are of global importance, such as 'Social Issues', including topics like '#BeatCancer', 'Swine Flu', '#Stoptheviolence' and 'Unemployment', the results indicate that the attribute, information roles (IRO) yields the best performance—since it is able to capture user interests via their information generation and consumption patterns.

From the results on utility measurement, we observe that for themes associated with current external events (e.g. 'Business-Finance', 'Politics', 'Entertainment-Culture' and 'Sports'), the attribute, activity behavior (ACT) yields high utility measures. This is because information diffusing in the network on current happenings, are often dependent upon the temporal pattern of activity of the users, i.e. their time of tweeting. For 'Human-Interest', 'Social Issues' and 'Hospitality-Recreation', we observe that the content creation attribute (CCR) yields the best performance in prediction, because it reveals the habitual properties of users in dissipating information on current happenings that they are interested in.

### 7.2.3 Comparative Study

Now we present a comparative study to evaluate the goodness of our proposed method—i.e. the DBN based social action model that is used to predict the users involved in diffusion at a future time slice. For the purpose of comparison, we utilize five different baseline methods. These methods are independently applied on the baseline and the attribute social graphs to predict the diffusion characteristics and the corresponding distortion measurements over time.

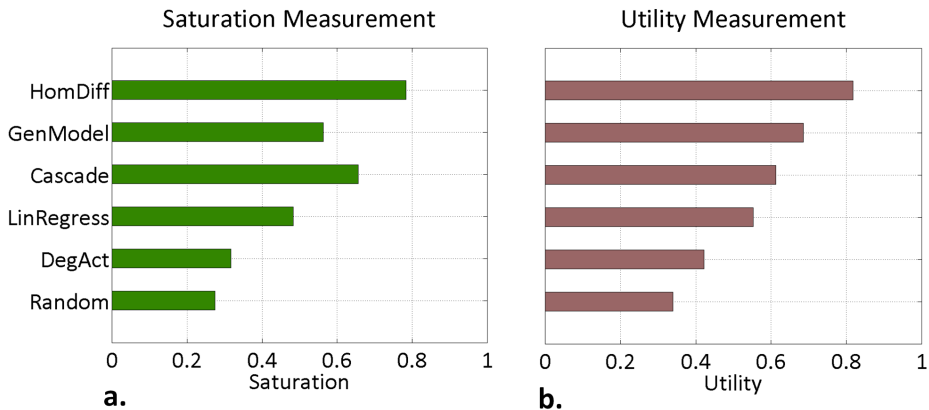

Figure 10: Comparison of mean saturation and utility measurement of predicted diffusion characteristics across different methods. Results are shown over the location attribute.

Our first baseline technique (`GenModel`) uses a Hidden Markov Model based generative framework of action prediction where tweeting (or not tweeting) are the observations, while the latent states correspond to the vulnerable and indifferent states (however unlike our method, the environmental features i.e. prior action, topic or friends' activity are not considered—hence this model is not context-aware). The next method (`Cascade`) is a threshold based model of 'global cascades' [34] based on the idea that a user participates in an action by changing her state to 'active' only when a certain sufficiently large fraction of her contacts have already done so. In the third baseline framework (`LinRegress`), we predict the probability of observed action (i.e. tweeting) using a linear regression model that uses the environmental features (i.e. prior action, topic and friends' activity) for training. The fourth baseline framework (`DegAct`) chooses users likely to participate in diffusion at a future time slice based on the degree of tweeting activity—the higher the activity measure at the current time slice, the higher is the probability of the user appearing in the diffusion series at the next time slice. Finally in the fifth baseline technique (`Random`), we select a set of users at random for the next time slice as participants in the diffusion process.

In Figure 10, we present the results of the comparative study over all these methods for a particular attribute location (LOC), averaged over all themes as well as time. In the case of both saturation measurement (computed over all diffusion characteristics, such as volume, reach, rate) and utility measurement (computed over search and news trends), we observe that our proposed method

Table 3: **Comparison of mean saturation and utility measurement of predicted diffusion (the higher, the better) across all methods and attribute and baseline social graphs.**

| METRIC | SOCIAL GRAPH | METHOD | | | | | |
|---|---|---|---|---|---|---|---|
| | | HomDiff | GenModel | Cascade | LinRgress | DegAct | Random |
| **Saturation Measurement** | **BAS** | **0.63** | **0.42** | **0.44** | **0.35** | **0.25** | **0.17** |
| | **LOC** | **0.78** | **0.56** | **0.65** | **0.48** | **0.31** | **0.27** |
| | IRO | 0.68 | 0.47 | 0.51 | 0.39 | 0.26 | 0.18 |
| | CCR | 0.67 | 0.45 | 0.49 | 0.36 | 0.28 | 0.21 |
| | ACT | 0.62 | 0.37 | 0.41 | 0.33 | 0.21 | 0.15 |
| **Utility Measurement** | BAS | 0.59 | 0.58 | 0.48 | 0.41 | 0.32 | 0.21 |
| | LOC | 0.51 | 0.52 | 0.44 | 0.31 | 0.28 | 0.16 |
| | IRO | 0.64 | 0.61 | 0.53 | 0.45 | 0.37 | 0.26 |
| | CCR | 0.72 | 0.65 | 0.57 | 0.52 | 0.39 | 0.29 |
| | **ACT** | **0.84** | **0.68** | **0.61** | **0.55** | **0.42** | **0.33** |

(`HomDiff`) yields the highest saturation in explaining the actual diffusion characteristics.

However, interestingly, in the case of saturation measurement, `Cascade` gives better performance compared to `GenModel`; while utility measurement reveals the opposite observation. We conjecture that saturation measurements, quantifying the user involvement, topological and temporal aspects of the diffusion process, are like to benefit more from methods that use the social graph's structural properties—hence `Cascade` performs better than `GenModel` in this case. While for utility measurements, correlation of diffusion characteristics with search and news trends are likely to benefit more from methods that utilize the users' tweeting behavior over time; hence the better performance from `GenModel`. Additionally, we note that for both cases, `Random` yields the worst performance, because it is not able to account for either the users' tweeting behavior or their network topology. Interestingly, note that `Cascade`, being able to utilize the social graph's structural properties, performs better than `GenModel` in case of saturation measurement.

Finally we summarize the performance of different methods across different attributes, as well as the baseline social graphs (averaged over all themes) in Table 3. We observe that our method (`HomDiff`) gives the highest saturation and utility measures compared to the baselines. Additionally, across the two measurement metrics, the attributes that yield best performance differ: being LOC and IRO for saturation measurement, while ACT and CCR for utility.

## 7.3 Discussion

We now discuss some of the implications of this work. Our extensive experimentation on the Twitter dataset has revealed that attribute homophily *indeed* impacts the diffusion process; however the particular attribute that can best explain the actual diffusion characteristics often depends upon: (1) the metric used to quantify diffusion, and the (2) topic under consideration. For example, the location attribute based homophily among users seems to predict well the topological diffusion characteristics (e.g. reach, spread) well, because users' local social neighborhoods are often clustered around commonality in their lo-cations. While in the case of topics, we have observed that activity behavior based homophily that can capture the temporal patterns of tweeting behavior of users can predict diffusion characteristics better for themes that are related to current external events, such as 'Politics', 'Technology-Internet' and 'Sports'. Note, these results certainly indicate that attribute homophily appears to impact the diffusion process characterization and we do provide empirical evidence of which kinds of themes and diffusion metrics seem to be more sensitive to specific attributes. Nevertheless, we have not provided a principled way of how to *learn* a particular attribute that would minimize the error in the predicted diffusion discovery. This avenue is left for future work.

We also acknowledge that our proposed method and experimentation is not without limitations. Results have indicated that for certain diffusion characteristics such as rate, our chosen set of attributes do not perform significantly better in predicting the diffusion process, compared to the case when we do not consider attribute homophily. This leaves future opportunities for us to explore presence of homophily along alternate attributes or attribute combinations that incorporate time, e.g. bursty or consistent tweeting behavior, or other temporal aspects of information dissipation [17].

Finally, it might be an obvious question to ask as to whether we can automatically infer the *optimal* attribute along which homophily exists in a network, and that best predicts the diffusion process. While we agree that automated homophily attribute selection is an important research topic, we believe that it was important to first show that there existed a relationship between diffusion and homophily, which had not been established in prior research. To elaborate more, it is important to note that our problem is different from a traditional feature/attribute selection based data mining problem. In a standard feature selection problem, the dependencies between the process and the features are established, so the goal is to select an optimal set of features. In this case, however, we are attempting to address a more fundamental question: i.e. whether there is any relationship at all between diffusion and attribute homophily. Note that this is the first work of its kind to investigate this dependency quantitatively.

Additionally, an important point of discussion is that in this article, to test our hypothesis on the relationship between diffusion and homophily, we have focused on the social media Twitter. This is because Twitter features a diverse user population who vary widely across attributes such as location and content creation behavior [24]; it also provides extensive evidence of information propagation (via tweets). However we acknowledge that our framework can be extended to other datasets with evidence of social actions.

It is also important to note that we have been interested in *strictly* studying how homophily with respect to a certain user attribute can lead to diffusion of content from a user $A$ to another user $B$; however, we refrain from making general claims about $A$ "socially influencing" $B$: because attitudes can become homophilous even without observable evidence of direct influence [8]. That is, the root cause of the observed auto-correlation [19] between the attribute homophily and the diffusion phenomenon can be either because of individuals getting social influenced (i.e. processes in which interactions with others causes individuals to conform e.g., people change their attitudes to be more similar to their friends), or because of a hidden condition or event, whose impact is correlated among instances that are closely situated in time or space [2].

# 8  Conclusions and Future Work

Classically, human communication activity involves mutual exchange of information, and the pretext of any social interaction among a set of individuals is a reflection of how our behavior, actions and knowledge can be modified, refined, shared or amplified based on the information that flows from one individual to another. Thus, over several decades, the structure of social groups, society in general and the relationships among individuals in these societies have been shaped to a great extent by the flow of information in them. Diffusion is hence the process by which a piece of information, an idea or an innovation flows through certain communication channels over time among the individuals in a social system.

Today, the pervasive use of online social media has made the cost involved in propagating a piece of information to a large audience extremely negligible, providing extensive evidences of large-scale social contagion. There are multi-faceted personal publishing modalities available to users today, where such large scale social contagion is prevalent: such as weblogs, social networking sites like MySpace and Facebook as well as microblogging tools such as Twitter. These communication tools are open to frequent widespread observation to millions of users, varied in diverse sets of attributes, and thus offer an inexpensive opportunity to capture large volumes of information flows at the individual level. If we want to understand the extent to which information spreads via these interactional affordances provided by different online social platforms, it is important to understand how the dynamics of propagation are likely to unfold within the underlying social media via the activity among the individuals, conditional over the "attribute similarities" among them: the extent to which people are likely to be affected by decisions of their friends and colleagues, or the extent to which "word-of-mouth" effects will take hold via interactional activity.

Our article investigates, for the first time, the relationship between information diffusion and homophily in online social media. To this effect, we have proposed a dynamic Bayesian network based framework to predict diffusion characteristics corresponding to different user attributes, such as location, activity behavior and information roles. We have also developed two kinds of metrics—saturation and utility measurement metrics that utilize the predicted characteristics to quantify the impact of attribute homophily in explaining the actual diffusion as well as external time series trends. Extensive experimentation on a Twitter dataset has revealed insights in favor of our hypothesis. Overall, attribute homophily is able to quantify the actual diffusion and external trends by a margin of $\sim 15-25\%$ lower distortion compared to cases when homophily is not considered. Comparison with baseline techniques has also indicated that our proposed method outperforms others in predicting diffusion characteristics subject to homophily, by $\sim 13-50\%$. To summarize, our conclusions include the following: (1)while diffusion has been studied in literature, we establish for the first time, that it is significantly affected by attribute homophily; (2)that this homophily-diffusion relationship is affected by the topic and diffusion metric.

There are several exciting directions to future work. In the future we are interested in investigating extensive sets of attributes on diverse social datasets. Additionally, instrumenting the interplay between homophily, emergent 'sync' of social actions and social influence in a network is also an exciting future direction. We are also interested in understanding the relationship between homophily and how "social influence" can spread via diffusion.

Diffusion paths in social networks, given observed homophily along a particular attribute, are also often associated with a temporal factor—either how frequently a topic diffuses in a network via communication, or how easily a network gets "infected" with a certain information. We have not addressed diffusion in these directions, however we have addressed a related temporal artifact; which is the diffusion property rate. As future research directions we would be interested to connect this measure of "delay" with models of network infection as well as the rate at which a user is responsible for diffusing a particular piece of information via her communication over an attribute or sets of attributes over which homophily exists.

Further, it would be interesting to focus on the temporal evolution of diffusion context given homophily on a certain attribute, possibly using a partially observable Markov decision process. The generic model of temporal evolution thus presents the scenario where the users engaged in tweeting activity are assumed to exhibit interactional behavior which are only partially observable to us e.g. a user who travels to different locations frequently, a user who have different schedules of activity exhibited online or a change in the real social relationship between a pair of users etc. The goal of such a model is then, to be able to infer the hidden contextual state related to the evolved social behavior over the chosen attribute, based on the observed interaction.

# References

[1] Lada Adamic and Eytan Adar. How to search a social network. *Social Networks*, 27(3):187–203, July 2005.

[2] Aris Anagnostopoulos, Ravi Kumar, and Mohammad Mahdian. Influence and correlation in social networks. In *KDD '08: Proceeding of the 14th ACM SIGKDD international conference on Knowledge discovery and data mining*, pages 7–15, 2008.

[3] Eytan Bakshy, Brian Karrer, and Lada A. Adamic. Social influence and the diffusion of user-created content. In *EC '09: Proceedings of the tenth ACM conference on Electronic commerce*, pages 325–334, New York, NY, USA, 2009. ACM.

[4] Frank M. Bass. A new product growth model for consumer durables. *Management Science*, 15:215–227, 1969.

[5] Peter S. Bearman, James Moody, and Katherine Stovel. Chains of affection: The structure of adolescent romantic and sexual networks. *American Journal of Sociology*, 110(1):44–91, July 2004.

[6] Michael J. Brzozowski, Tad Hogg, and Gabor Szabo. Friends and foes: ideological social networking. In *CHI '08: Proceeding of the twenty-sixth annual SIGCHI conference on Human factors in computing systems*, pages 817–820, New York, NY, USA, 2008. ACM.

[7] Ronald S. Burt. Toward a structural theory of action: Network models of social structure, perception and action. *The American Journal of Sociology*, 90(6):1336–1338, 1982.

[8] Robert B. Cialdini and Noah J. Goldstein. Social influence: Compliance and conformity. *Annual Review of Psychology*, 55:591–621, February 2004.

[9] David Crandall, Dan Cosley, Daniel Huttenlocher, Jon Kleinberg, and Siddharth Suri. Feedback effects between similarity and social influence in online communities. In *KDD '08: Proceeding of the 14th ACM SIGKDD international conference on Knowledge discovery and data mining*, pages 160–168, 2008.

[10] Munmun De Choudhury, Yu-Ru Lin, Hari Sundaram, Kasim Selcuk Candan, Lexing Xie, and Aisling Kelliher. How does the data sampling strategy impact the discovery of information diffusion in social media? In *ICWSM '10: Proceedings of the Third International Conference on Weblogs and Social Media*, San Jose, CA, May 2010. AAAI Press.

[11] Jager W. Delre, S.A. and M.A. Janssen. Diffusion dynamics in small-world networks with heterogeneous consumers. *Comput. Math. Organ. Theory*, 12(2):185–202, 2007.

[12] Scott L. Feld. The focused organization of social ties. *American Journal of Sociology*, 86(5):1015–1035, 1981.

[13] Andrew T. Fiore and Judith S. Donath. Homophily in online dating: when do you like someone like yourself? In *CHI '05: CHI '05 extended abstracts on Human factors in computing systems*, pages 1371–1374, 2005.

[14] Lind P.G. Gonzalez, M.C. and H.J. Herrmann. Model of mobile agents for sexual interaction networks. *European Physics Journal B*, 49:371–376, 2006.

[15] Daniel Gruhl, R. Guha, David Liben-Nowell, and Andrew Tomkins. Information diffusion through blogspace. In *WWW '04: Proceedings of the 13th international conference on World Wide Web*, pages 491–501, New York, NY, USA, 2004. ACM.

[16] David Kempe, Jon Kleinberg, and Éva Tardos. Maximizing the spread of influence through a social network. In *KDD '03: Proceedings of the ninth ACM SIGKDD international conference on Knowledge discovery and data mining*, pages 137–146, 2003.

[17] Gueorgi Kossinets, Jon Kleinberg, and Duncan J. Watts. The structure of information pathways in a social communication network. In *KDD '08: Proceeding of the 14th ACM SIGKDD international conference on Knowledge discovery and data mining*, pages 435–443, New York, NY, USA, 2008. ACM.

[18] Gueorgi Kossinets and Duncan J. Watts. Origins of homophily in an evolving social network. *American Journal of Sociology*, 115(2):405–450, September 2009.

[19] Timothy LaFond and Jennifer Neville. Randomization tests for distinguishing social influence and homophily effects. In *WWW '10: Proceedings of the International World Wide Web Conference*, pages 601–610, New York, NY, USA, April 2010. ACM.

[20] Jure Leskovec, Lars Backstrom, and Jon Kleinberg. Meme-tracking and the dynamics of the news cycle. In *KDD '09: Proceedings of the 15th ACM SIGKDD*, pages 497–506, 2009.

[21] D. Liben-Nowell and Jon Kleiberg. Tracing information flow on a global scale using internet chain-letter data. *PNAS*, 105(12):4633–4638, 2008.

[22] Miller Mcpherson, Lynn S. Lovin, and James M. Cook. Birds of a feather: Homophily in social networks. *Annual Review of Sociology*, 27(1):415–444, 2001.

[23] Miller McPherson and Lynn Smith-Lovin. Homophily in voluntary organizations: Status distance and the composition of face-to-face groups. *American sociological review*, 52(3):370–379, 1987.

[24] Chih-Hui Lai Mor Naaman, Jeffrey Boase. Is it really about me? message content in social awareness streams. In *CSCW '10: Proceedings of the 2010 ACM Conference on Computer Supported Cooperative Work*, New York, NY, USA, 2010. ACM.

[25] Kevin Murphy. *Dynamic Bayesian Networks: Representation, Inference and Learning.* PhD thesis, UC Berkeley, Computer Science Division, July 2002.

[26] M. E. Newman. Scientific collaboration networks. ii. shortest paths, weighted networks, and centrality. *Phys Rev E Stat Nonlin Soft Matter Phys*, 64(1 Pt 2), July 2001.

[27] Lawrence R. Rabiner. A tutorial on hidden markov models and selected applications in speech recognition. pages 267–296, 1990.

[28] Nakano R. Saito, K. and M. Kimura. Prediction of information diffusion probabilities for independent cascade model. *Knowledge-based Intelligent Information and Engineering Systems*, 3:67–75, 2008.

[29] Xiaodan Song, Belle L. Tseng, Ching-Yung Lin, and Ming-Ting Sun. Personalized recommendation driven by information flow. In *SIGIR '06: Proceedings of the 29th annual international ACM SIGIR conference on Research and development in information retrieval*, pages 509–516, 2006.

[30] Avaré Stewart, Ling Chen, Raluca Paiu, and Wolfgang Nejdl. Discovering information diffusion paths from blogosphere for online advertising. In *ADKDD '07: Proceedings of the 1st international workshop on Data mining and audience intelligence for advertising*, pages 46–54, New York, NY, USA, 2007. ACM.

[31] Eric Sun, Itamar Rosenn, Cameron Marlow, and Thomas Lento. Gesundheit! modeling contagion through facebook news feed. In *Proceedings of the Third International Conference on Weblogs and Social Media*, San Jose, CA, May 2009.

[32] Jie Tang, Jimeng Sun, Chi Wang, and Zi Yang. Social influence analysis in large-scale networks. In *KDD '09: Proceedings of the 15th ACM SIGKDD international conference on Knowledge discovery and data mining*, pages 807–816, New York, NY, USA, 2009. ACM.

[33] Xiaojun Wan and Jianwu Yang. Learning information diffusion process on the web. In *WWW '07: Proceedings of the 16th international conference on World Wide Web*, pages 1173–1174, New York, NY, USA, 2007. ACM.

[34] D. J. Watts. A simple model of global cascades on random networks. *PNAS*, 99:5766–5771, 2002.

[35] D. J. Watts, P. S. Dodds, and M. E. J. Newman. Identity and search in social networks. *Science*, 296(5571):1302 – 1305, May 2002.